\documentclass[amssymb,aps,superscriptaddress,pre,twocolumn,floatfix]{revtex4-1}
\usepackage[]{amsmath}
\usepackage{graphicx}
\usepackage{epstopdf}
\usepackage{dcolumn}
\usepackage[usenames, dvipsnames]{color}
\usepackage{xcolor,colortbl}
\usepackage{array}
\usepackage{bigstrut}

\newcommand{\rr}{\raggedright}
\newcommand{\tn}{\tabularnewline}

\definecolor{Gray}{gray}{0.85}

\definecolor{Red}{rgb}{1,0.8,0.8}
\definecolor{Orange}{rgb}{1,1,0.85}
\definecolor{Green}{rgb}{0.8,1,0.8}
\definecolor{LightCyan}{rgb}{0.8,1,1}
\definecolor{Cyan}{rgb}{0.8,1,1}
\definecolor{White}{gray}{1}
\newcommand{\cR}{\cellcolor{Red}}
\newcommand{\cO}{\cellcolor{Orange}}
\newcommand{\cG}{\cellcolor{Green}}
\newcommand{\cC}{\cellcolor{Cyan}}
\newcommand{\cW}{\cellcolor{White}}

\newcolumntype{a}{>{\columncolor{Gray}}c}
\newcolumntype{b}{>{\columncolor{white}}c}

\definecolor{DarkCyan}{cmyk}{1.0,0.0,0.0, 0.4}
\definecolor{LightRed}{cmyk}{0,0.8,0.9, 0}

\begin{document}

\title{Evidence of Herding and Stubbornness in Jury Deliberations}

\author{Keith Burghardt}
\email[]{kaburghardt@ucdavis.edu}
\affiliation{Dept. of Computer Science, University of California at Davis, Davis, California, USA, 95616}
\affiliation{Dept. of Political Science, University of California at Davis, Davis, California, USA, 95616}

\author{William Rand}
\affiliation{Poole College of Management, North Carolina State University, Raleigh, North Carolina, USA, 27695}

\author{Michelle Girvan}
\affiliation{Dept. Of Physics, University Of Maryland, College Park, Maryland, USA, 20742}
\affiliation{Institute for Physical Science and Technology, University Of Maryland, College Park, Maryland, USA, 20742}
\affiliation{Santa Fe Institute, Santa Fe, New Mexico USA, 87501}

\date{\today}

\begin{abstract}

We explore how the mechanics of collective decision-making, especially of jury deliberation, can be inferred from macroscopic statistics. We first hypothesize that the dynamics of competing opinions can leave a ``fingerprint'' in the joint distribution of final votes and time to reach a decision. We probe this hypothesis by modeling jury datasets from different states collected in different years and identifying which of the models best explains opinion dynamics in juries. In our best-fit model, individual jurors have a ``herding'' tendency to adopt the majority opinion of the jury, but as the amount of time they have held their current opinion increases, so too does their resistance to changing their opinion (what we call ``increasing stubbornness''). By contrast, other models without increasing stubbornness, or without herding, create poorer fits to data. Our findings suggest that both stubbornness and herding play an important role in collective decision-making. 

\end{abstract}




\maketitle


What mechanisms underlie collective decision-making? Recent research into opinion dynamics has compared statistical patterns in empirical data to models of opinion dynamics \cite{Facebook,ComplexContagion,LermanCC,CorrelationData,PollVolatility,VoterScaling1,VoterScaling2,CombineModels} and tested how opinions change in controlled experimental settings \cite{ModelingRealOpDyn,SocialInfluenceCrowd}. Although both methods have provided substantial insight extending the early studies of collective opinion dynamics \cite{VM1,VM2}, they have generally not tackled one question our paper aims to answer: how do non-consensus decisions occur \cite{Zealot2}? For example, is it possible to distinguish influence from non-influence when we observe group opinions \cite{InfluenceVsHomophily,HomophilyAndContagion}? In other words, if members of a group share the same opinion, can we tell if they arrived at that opinion independently or as a result of their interactions (e.g., through a tendency to ``follow'' or imitate one another)? If this kind of influence does plays a role, why do opinions often remain in non-consensus? 

In this paper, we probe these questions by studying jury deliberation. Our work compliments previous research where judicial rulings were found to be affected by factors unrelated to the specific cases \cite{JudgeRuling}. By analyzing jury data in bulk, we aim to understand how the mechanisms of jury opinion formation not directly related to the facts of the case (such as influence and stubbornness) couple together with the laws defining hung juries (situations in which jury opinions are considered too divided to reach a verdict) to shape decision-making patterns observed in data. Furthermore, while groups are thought to create better decisions than single individuals \cite{JuryThm,Galton1908,WikiWisdomOfCrowds,SocialInfluenceCrowd}, a recent model suggests that correlated juror decisions undermine their collective accuracy \cite{JuryThmCorrel}, a conclusion supported by experiments on crowd wisdom \cite{SocialInfluenceCrowd}. We have found that there are significant correlations between decisions (see Fig. \ref{fig:VoteCorrelation}), which, together with the previous model \cite{JuryThmCorrel}, may help explain why juries appear to perform \emph{worse} than individual judges at correctly acquitting defendants \cite{Spencer2007}. We are therefore motivated to understand the mechanisms that drive these correlations, which may point to ways in which the quality of jury decisions can be improved.

\section*{Features of the Data}

Jury deliberation is an ideal test bed for models of opinion dynamics. Jurors are exposed to the same information during the trial, are instructed not to discuss the trial with non-jurors, and cannot learn about the trial from outside sources  \cite{CrimJuryInstructionsCA,CivilJuryInstructionsCA,ORJuryInstructions,WAJuryInstructions,NEJuryInstructions}, therefore opinion variation between jurors is likely due to internal factors, such as influence instead of external factors, such as varying levels of information. 

\begin{figure}[t]
	\centering
	\includegraphics[width=0.99\columnwidth]{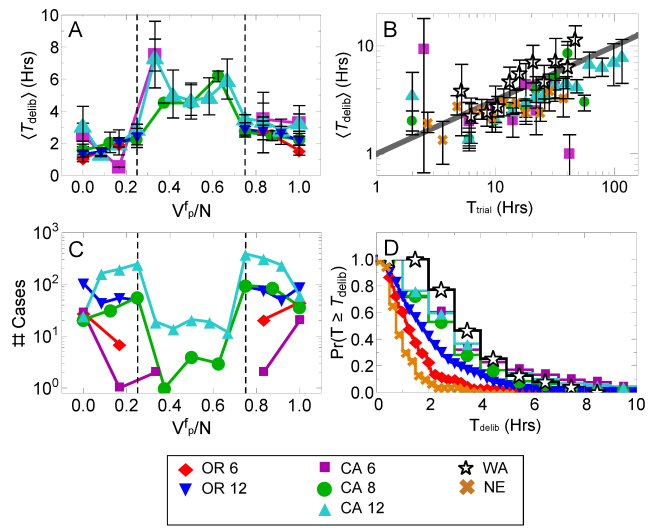}
	\caption{\label{fig:JuryData}{Data attributes.} (a) The mean deliberation time peaks when the $V^\text{f}_\text{p}/N\approx0.3$ and $0.6$, where $V^\text{f}_\text{p}/N$ is the fraction of jurors voting for the plaintiff. (b) The mean deliberation time scales as $(T_\text{trial})^{1/2}$ across several datasets, even though $T_\text{trial}$ correlates only weakly with $V^\text{f}_\text{p}/N$ (Fig. \ref{fig:TrialTimeVsFinalVote}). (c) The distribution of $V^\text{f}_\text{p}/N$ shows peaks when $V^\text{f}_\text{p}/N\ge \phi$ or $V^\text{f}_\text{p}/N\le 1-\phi$, where $\phi=0.75$, the thresholds between which juries hang  \cite{ORJuryData,CAHungJuries}. (d) The complimentary cumulative distribution of deliberation times is heavy-tailed across datasets. Data is taken from Oregon \cite{ORJuryData}, California \cite{CAJuryData}, Washington, and Nebraska \cite{USJuryPaper1,USJuryDataPaper2}, and error bars represent 90\% confidence intervals in the mean.
	}
\end{figure}

 In this paper, we model jury datasets for civil trials in Oregon (OR) \cite{ORJuryData} and California (CA) \cite{CAJuryData}. We find qualitatively similar behavior in the Washington (WA) and Nebraska (NE) datasets whose data is less complete \cite{USJuryPaper1,USJuryDataPaper2} (Fig.~\ref{fig:JuryData}), and for criminal trials in the OR dataset (see Fig. \ref{fig:ORAttributes}). We split data by the number of jurors in each case. The OR 6 dataset, for example, corresponds to the Oregon civil trials with six jurors.

The purpose of our modeling efforts is to explain four features of the data shown in Fig. \ref{fig:JuryData}.  First, we aim to model why the mean deliberation time, $\langle T_\text{delib}\rangle$, is lowest when the fraction of jurors voting for the plaintiff in the final vote, $V^\text{f}_\text{p}/N$ is $0$ or $1$, and highest when $V^\text{f}_\text{p}/N \approx 0.3$ or $V^\text{f}_\text{p}/N\approx0.6$ (Fig. \ref{fig:JuryData}a). For each trial, we look at the break down between for-plaintiff and for-defendant votes. We also normalize the number of jurors voting for the plaintiff in the final vote, $V^\text{f}_\text{p}$, by the jury size, $N$. Juries are dismissed and a new trial is given to the defendant in a civil trial if $1-\phi < V^\text{f}_\text{p}/N < \phi$, where $\phi$ is $0.75$ for the OR and CA civil trials \cite{ORJuryData,CAHungJuries}. This property, observed across all $N$, also helps explain why juries tend to reach supermajorities, in which $V^\text{f}_\text{p}/N \ge \phi$ or $V^\text{f}_\text{p}/N\le 1-\phi$ (Fig. \ref{fig:JuryData}c). We also observe that $\langle T_\text{delib}\rangle$ scales with the trial time, $T_\text{trial}$, as $\langle T_\text{delib}\rangle\sim(T_\text{trial})^{1/2}$ (Fig. \ref{fig:JuryData}b), a property not strongly correlated with the final vote (Fig. \ref{fig:TrialTimeVsFinalVote}), even though both affect the mean deliberation time. Interestingly, the mean deliberation time does not scale strongly with the number of jurors, which is counter to many intuitive models (Fig.~\ref{fig:TdelibVsN}) \cite{VMConsensus,MajorityRule}.  Finally, in Fig. \ref{fig:JuryData}d, we notice that 
deliberation time distribution is heavy-tailed.

In this paper, we fit empirical joint distributions of final votes and deliberation times to model distributions through maximum likelihood estimation of model parameters, therefore, counter-intuitively, we infer opinion \emph{dynamics} without access to time-series data. Recent work, however, has shown that different dynamical models of group opinion formation create different distributions in the time for groups to reach consensus \cite{VMMaj,VMDist}, which inspires us to reverse engineer the model that best matches the jury data. By matching the joint distribution instead of either distribution alone strongly limits the possible dynamical models that can explain the data. For example, in contrast to many models of group decision-making, juries rarely reach complete agreement before they stop deliberating, therefore it may be possible to match the deliberation time alone with unrealistic models, in which all jurors reach agreement. We do not claim, however, that the models we present are the only possible models that can match the data well, but are instead simple models that illustrate mechanisms that drive opinion dynamics.

A major limitation in the data, however, is that no two trials are exactly the same, therefore aggregating over heterogeneous trials may strongly affect our results \cite{SimpsonEffect}. 
To test for this effect, we split data into more homogenous groups with the same $N$ and similar $T_\text{trial}$, because $N$ affects $V^\text{f}_\text{p}$ and $T_\text{trial}$ affects $\langle T_\text{delib}\rangle$. We find, however, that splitting the data does not affect the qualitative behavior of $V^\text{f}_\text{p}/N$, and $T_\text{delib}$ (E.g., see Fig.~\ref{fig:JuryData}). These results so far suggest that heterogeneity should not significantly affects our findings.

We next introduce three null models, in which jurors reach their opinions independently, then six models incorporating influence, though herding behavior and individual stubbornness, and discuss how well each matches the data. 

\section*{Null Models}

\begin{figure}[t]
	\hspace{-12pt}
	\centering
	\includegraphics[width=0.99\columnwidth]{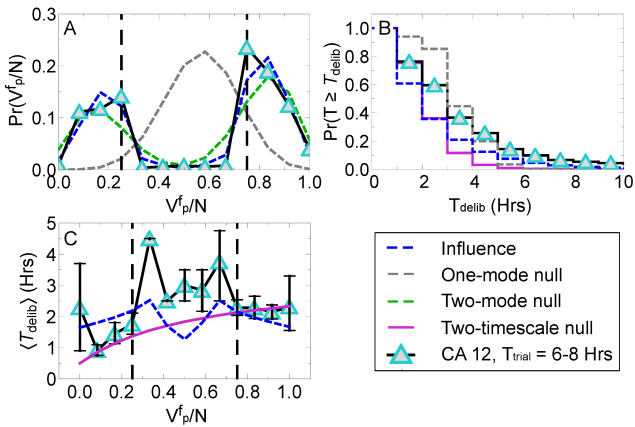}
	\vspace{-1em}
	\caption{
Example fits of the influence model compared to null models: one- and two-mode models where each juror's vote is reshuffled within a set of votes, and the two-timescale model. (a) $\textrm{Pr}(V^\text{f}_\text{p}/N)$ versus $V^\text{f}_\text{p}/N$, (b) $\textrm{Pr}(T\ge T_\text{delib})$ versus $T_\text{delib}$, and (c) $\langle T_\text{delib}\rangle$ versus $V^\text{f}_\text{p}/N$.
	}
	\label{fig:JuryModels}
\end{figure}

We first create an independent, random vote null model with which to compare other models. For this first model, for each dataset, we reshuffle all juror votes, which creates a binomial distribution of final votes. Not surprisingly, this ``one-mode null model'' fits data poorly; therefore we propose a slightly more nuanced ``two-mode null model.'' For this, we split the jury data into those with majority for-plaintiff final votes ($V^\text{f}_\text{p}/N > 0.5$) and the rest ($V^\text{f}_\text{p}/N \le 0.5$), reshuffle juror votes of each subset separately, and then combine the distributions. In both cases, we fix $P(T_\text{delib}|V^\text{f}_\text{p}/N)$, the conditional probability for juries to stop deliberating at time $T_\text{delib}$, given the fraction of for-plaintiff votes in the final vote, $V^\text{f}_\text{p}/N$, to exactly match the empirical data, as an unrealistic but best-case scenario of these null models.
Both models produce poor fits of the data compared to other models (Fig. \ref{fig:JuryModels}e, \ref{fig:ModelComparisons}, \& \ref{fig:ORAttributes}), with the exception of CA 12 ($T_\text{trial} = 34-61$ hours) in which the two-mode model fits data better than any influence model. Overall, however, a simple model in which opinions are picked at random, independently of each other, does not provide a compelling explanation of the data.

We also create a ``two-timescale'' null model of the deliberation time distribution, in which the time for each juror to make their pre-determined final decision is independent (exponentially distributed), but depends on whether their decision is for the plaintiff or not (hence ``two-timescale''). Deliberation ends when the last juror makes their final decision. Separate fitting parameters are used for for-plaintiff and for-defendant votes because for-plaintiff votes usually take longer than for-defendant ones (p-value $<2\times10^{-2}$ based on the Mann-Whitney U test for CA 6, CA12, OR 6, and OR 12 datasets, no significant difference for the CA 8 dataset), and it allows for this null model to better agree with the data. We determined distributions by Monte Carlo sampling $10^5$ times for each $V^\text{f}_\text{p}$ such that $P(V^\text{f}_\text{p}/N)$ is fixed to be the empirical data distribution as a best-case scenario. In this way, the two-timescale null model is meant to explain how juries stop deliberating, not how they reach their final vote. We find, however, that this model creates a poorer fit to the observed data than the full influence model (to be discussed shortly), despite artificially fixing $P(V^\text{f}_\text{p}/N)$. While other plausible time distributions could be used and the assumption of a homogeneous distribution might not be ideal, disagreement between this idealized model and data point to limitations in similar null models.

\section*{Influence Models}

\subsection*{Full model} 
\begin{figure}[t]
	\hspace{-12pt}
	\centering
	\includegraphics[width=0.99\columnwidth]{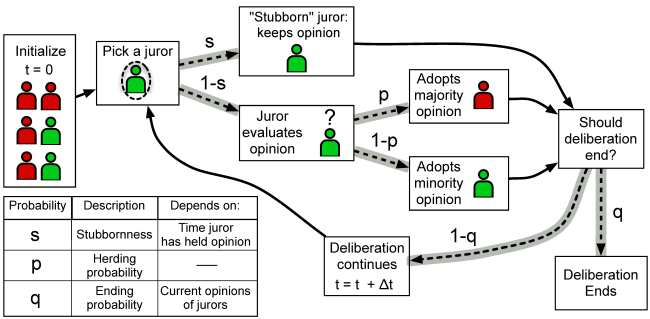}
	\caption{Schematic of the influence model. Solid lines correspond to deterministic transitions, while dashed lines correspond to probabilistic transitions. Jurors are first initialized to have one of two opinions (for the plaintiff or defendant). At each timestep, a random juror is picked and considers re-evaluating their opinion with probability $1-s$, where $s$ is ``stubbornness''. If they do re-evaluate, they pick the majority opinion with probability $p$, and the minority with probability $1-p$. At the end of each timestep, the jury stops deliberating with probability $q$. See Eq.~\eqref{eq:s},~\eqref{eq:mu}, \& \eqref{eq:q} for definitions of $s$ and $q$.
}\label{fig:ModelSchematic}
\end{figure}

Given the relatively poor performance of the null models, we propose an ``influence with increasing stubbornness'' model that can better describe the datasets. Within the large space of plausible models, we focus on a simple model with few parameters, and then check whether any of these parameters could be removed without affecting the quality of the fit. Furthermore, we focus on a model with herding because of the correlations between juror opinions seen in the data (Fig. \ref{fig:VoteCorrelation}), which may suggest that jurors have a tendency to follow the majority opinion. In the model, jurors tend to adopt the majority opinion and juries end deliberation at a rate that depends on the current vote (number of jurors currently leaning for the plaintiff and for the defendant). The former incorporates a simple mechanism for juror influence that enables the supermajorities observed in data, while the latter captures jury resistance to hung conditions. In addition, we add a stubbornness property, in which jurors increasingly hold on to their current opinion. This facilitates the strong non-consensus patterns from data. More specifically, as shown in Fig.~\ref{fig:ModelSchematic}, at each timestep in the model (where a timestep is chosen to be 1 minute, see Methods), a random juror is selected and considers re-evaluating their current opinion with probability $1-s$, where $s$ reflects their stubbornness and depends on how long they've held their current opinion. If they do re-evaluate, they pick the majority opinion with probability $p$, and the minority with probability $1-p$. At the end of each timestep, the jury stops deliberating with probability $q$, which depends on the current set of juror opinions. 

The stubbornness probability $s$, depends both on how long the juror has held their current opinion and whether the current set of opinions meet the hung condition:
\begin{equation}
\label{eq:s}
s =\sum_{i=0}^{\tau/\Delta t} \mu_\text{eff}(t_0+i \Delta t) \Delta t,
\end{equation}
\noindent
where $t_0$ is the time a juror adopted its most recent opinion, $\tau$ is the time a juror has held their current opinion, $\Delta t$ is the length of a simulation time step, and $\mu_\text{eff}(t)$ is the rate jurors become more stubborn:
\begin{equation}
\mu_\text{eff} (t)= 
\begin{cases}
\mu & \frac{V_\text{p}(t)}{N}\le1-\phi,~\frac{V_\text{p}(t)}{N}\ge \phi \\
f \mu & 1-\phi<\frac{V_\text{p}(t)}{N}<\phi 
\end{cases},
\label{eq:mu}
 \end{equation}
where $f$ is the reduction in this rate when juries are hung (the current vote, $V_\text{p}(t)$, divided by $N$ is between the jury hanging thresholds $1-\phi$ and $\phi$). If we set the stubbornness probability $s$ to a constant, that would only have the general affect of changing the timescale of the dynamics. We incorporate \emph{increasing} stubbornness ($s$ grows with $\tau$) as a behavioral hypothesis, which has previously been shown to help explain voter behavior in elections \cite{FastConsensus,CombineModels}. The jury's tendency to reach a non-hung decision is captured by making the stubbornness rate $\mu_\text{eff}(t)$ lower under hung conditions, meaning that jurors 
do not hold onto their opinions as strongly as they would otherwise, presumably to lessen the probability that the jury hangs.

At the end of each timestep, the probability for a jury to stop deliberating, $q$, is determined:
\begin{equation}
q = \begin{cases}
q_0 + \alpha |\frac{V_\text{p}(t)}{N} -1/2| & \frac{V_\text{p}(t)}{N}\le1-\phi,~\frac{V_\text{p}(t)}{N}\ge \phi \\
q_0 & 1-\phi<\frac{V_\text{p}(t)}{N}<\phi
\end{cases},
\label{eq:q}
\end{equation}
where $q_0$ is the base rate of quitting, and $\alpha |\frac{V_\text{p}(t)}{N} -1/2| $ is expected to be greater than 0 because the jury is more likely to stop deliberating if it is not currently hung. If they do stop deliberating at time $t$, then the time and the final vote, $V^\text{f}_\text{p}=V_\text{p}(T_\text{delib}=t)$, are recorded.

To summarize, the influence model, as shown in Fig. \ref{fig:ModelSchematic}, involves three different transition probabilities: $p$, $q$, and $s$. These transition probabilities are constructed from a total of four fitting parameters: $\mu$, $\alpha$, $f$, and $p$, and three fixed parameters: $b$, the bias of the initial condition; $\Delta t$, the length of a time step; and $q_0$, which are discussed further in Methods. Fig. \ref{fig:JuryModels}) shows that not only can the model explain vote and time distributions, but it can also explain the peaks in deliberation time near the critical fraction of voters $V^\text{f}_\text{p}/N \approx 0.3$ and $0.6$. This appears to be due to important factors included in the influence model: the instability of juries having 50/50 split decisions, and the ability for juries to stop deliberating even then they have not reached complete consensus (see Supporting Information).

\begin{figure}[t]
	\hspace{-12pt}
	\centering
	\includegraphics[width=0.99\columnwidth]{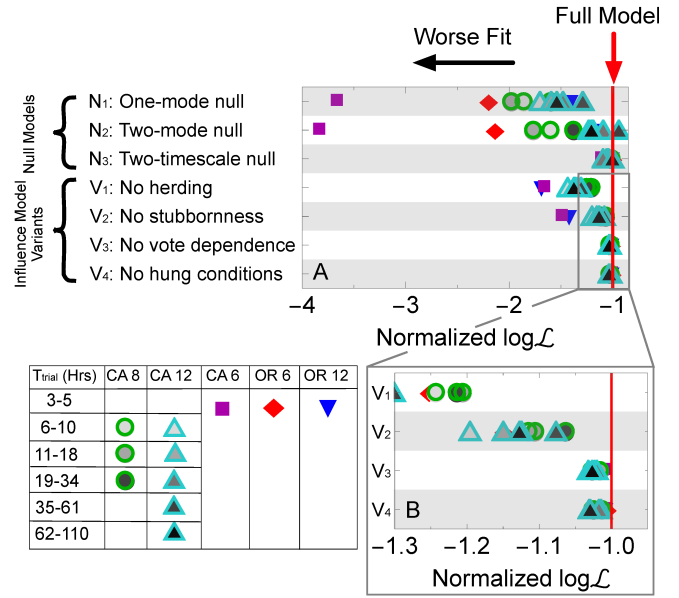}
\caption{\label{fig:ModelComparisons} {{Comparison of Models.}} Normalized log-likelihood functions for the null models and the influence model variants to illustrate comparison with the full influence model. For each dataset indicated in the legend, log-likelihood functions for these models were normalized by $|\text{log}(\mathcal{L}_\textrm{full})|$, the log-likelihood function of the full model, therefore models above -1 explain the data better than the full influence model, while those below -1 perform worse. (a) The relative fit of the one-mode, two-mode, and two-timescale null models, along with  ``no herding'' model, in which $p=0.5$, ``no stubbornness'' model with $\mu=0$, ``no vote dependence'' model, in which the model dynamics do not depend on the number of jurors voting for the plaintiff, and the ``no hung conditions'' model, in which jury dynamics do not depend on whether the jury is currently hung. (b) In a zoomed-in graph, the influence model variants seen in (a) perform worse than the full model.
}
\end{figure}

\subsection*{Variations of the full model} Having developed a model that explains the data better than the simple null models, we construct variants of the full model in order to identify which mechanisms are most important for capturing the observed patterns. First, we test whether herding affects jury trials by setting $p=0.5$ (Figs.~\ref{fig:ModelComparisons} \& \ref{fig:ORAttributes}). If $p=0.5$, a juror would have equal preference to pick the majority opinion as the minority one. We see that the fit is significantly worse, therefore herding appears to affect the outcomes of jury trials. We next test the role of increasing stubbornness by setting $\mu_\text{eff}=0$. Removing the increasing stubbornness parameter, however, produces significantly poorer fits to the data (Figs.~\ref{fig:ModelComparisons} \& \ref{fig:ORAttributes}). A similar conclusion is reached in previous work that matches a model to election data in several European countries \cite{CombineModels}. Because highly disparate datasets have similar conclusions about the importance of increasing stubbornness, we believe it plays a fundamental role in opinion dynamics. Setting the stubbornness probability $s$ to a constant greater than 0 should only generally decrease the timescale of the dynamics, presumably making the final vote distribution more similar to the initial vote distribution, therefore in the interest of space, we leave out further model variants of this type. Finally, to better understand how the hung conditions affect jury behavior, we fit a model with no dependence on hanging: $\mu_\text{eff}(t)=\mu$ and $q(t)=q_0 + \alpha |V_\text{p}(t)/N -1/2|$. In this ``no hung conditions'' variant, neither stubbornness rate, nor the quitting rate, depends on whether the jury is currently hung. The probability for the jury to end deliberations, however, still increases linearly with the amount of consensus among jurors. To test the importance of the current vote has on jury dynamics, we create a ``no vote dependence'' variant in which $\mu_\text{eff}(t)=\mu$, and $q$ is a fitted constant. Both of these variants show poorer agreement to the data compared to the full model (Figs.~\ref{fig:ModelComparisons} \& \ref{fig:ORFitComparison}). We finally tested removal of the hung conditions from either the stubbornness rate (Eq.~\eqref{eq:mu}) or the stopping probability (Eq.~\eqref{eq:q}), but not both. We find that removing the hanging dependence of the stubbornness rate fits the data worse than removing the hanging dependence of the stopping probability (Figs.~\ref{fig:NoHungComparison} \& \ref{fig:ORNoHungComparison}). Hanging may therefore affect how juror opinions change more than it affects how juries decide to end deliberations.

In summary, the full model agrees with data significantly better than the null models: one-mode null model, two-mode null model, and two-timescale null, as well as variants that remove herding, stubbornness, hung-conditions, and vote-dependent behavior.

\begin{table*}[t]
{\centering
Table 1: Model Parameter Descriptions and Fits
}
\begin{flushleft}
\label{tab:ModelFits}
\vspace{0em}
\begin{minipage}{.98\textwidth}
\begin{minipage}{0.15\textwidth}\centering
{\renewcommand{\arraystretch}{1.4}
\begin{center}

{\small
\resizebox{\textwidth}{!}{
\begin{tabular}{ | r p{1.9cm} |}
    \hline 
\multicolumn{2}{|c|}{ \bf Parameter} \\ [-0.6em]
\multicolumn{2}{|c|}{ \bf Descriptions } \\  \hline

   \cR  $p$&  \cR \rr Herding probability \tn \hline
    \cO $\mu$  & \cO \rr Stubbornness rate \tn \hline
    \cG $f$ & \cG \rr  Reduces $\mu$ when hung \tn \hline 
    \cC $\alpha$ & \cC \rr Controls rate of ending  deliberation \tn  
\hline
\end{tabular} }}
\end{center}}
\end{minipage}
\begin{minipage}{.83\textwidth}\centering
{\renewcommand{\arraystretch}{0.95}
\setlength\tabcolsep{1pt}
\resizebox{\textwidth}{!}{
\begin{tabular}{|>{\columncolor{Gray}}r| c| c| >{\columncolor{Red}}c| >{\columncolor{Red}}c| >{\columncolor{Orange}}c|>{\columncolor{Orange}} c| >{\columncolor{Green}} c|>{\columncolor{Green}}  c|>{\columncolor{LightCyan}} c|>{\columncolor{LightCyan}}  c|}
  \hline
  \multicolumn{11}{|c|} {\cW \bf Parameter Fits \bigstrut} \\ \hline 
  Data & $T_\text{trial}$ (Hrs) & Num. Trials & $\hat{p}$ & 90\% Conf. & $\hat{\alpha}$ ($\text{min}^{-1}$)& 90\% Conf. & $\hat{\mu}$ ($\text{min}^{-1}$) & 90\% Conf. & $\hat{f}$ & 90\% Conf. \\
  \hline 
\vspace{-7pt}
&&&&&&&&&&\\
OR 6 &--   	     &	101&0.956 & 0.94-0.98  & 0.0163 & 0.014-0.020& 0.17& 0.09-0.33  & 0.037& 0.0-0.08 \\
OR 12 & 	 -- &560&0.873 & 0.85-0.89 & 0.012 & 0.010-0.014&0.070 &0.05-0.09 & 0.028 & 0.0-0.09 \\
  CA 6    & --       & 53&    0.975  &0.97-0.98& 0.0080 &0.007-0.009&  0.042     &0.03-0.07&1.0&0.9-1.0\\
  CA 8    & 6-10  &171&  0.913   &0.85-0.95& 0.0154& 0.014-0.017    & 0.37    &0.2-0.7     &0.11        &0.0-0.2   \\ 
  CA 8    & 11-18& 121&  0.858  & 0.80-1.0   &0.0104&0.010-0.012     & 0.49   &0.2-1.0     &0.24      &0.0-0.6  \\
  CA 8    & 19-34&32 &   0.775  &0.75-0.80&0.0065 & 0.005-0.008&  0.50       &0.40-0.67&0.011&0.0-0.02\\
  CA 12  & 6-10  &502 &  0.864 & 0.83-0.91& 0.0146&0.014-0.016&0.63 & 0.4-1.0 & 0.22& 0.1-0.4 \\
  CA 12  & 11-18& 656&  0.811 &0.78-0.86 &0.011&0.010-0.013 &0.38 & 0.2-0.5&0.25     &0.2-0.4 \\
  CA 12  & 19-34&402&  0.812 &0.80-0.85  &0.0079&0.007-0.009&0.51 & 0.4-0.6& 0.19 &0.1-0.4\\
  CA 12  & 35-61& 111&  0.747 &0.71-0.80 &0.0068 &0.005-0.007 &0.28 & 0.15-0.67&0.13 &0.1-0.2 \\
  CA 12  & 62-110&42&0.765    &0.75-0.78&0.0040   &0.0035-0.0045&   0.33    &0.2-1.0& 0.001&0.0-0.015\\\hline
\end{tabular}
}}
\end{minipage}
\end{minipage}
\end{flushleft}
\end{table*}

\subsection*{Findings}

What does the influence model suggest about jury deliberation? To begin to answer this question, we examine the best-fit model parameters for the different datasets (Tab.~1). Similar results are found when we look at criminal data from Oregon as well (Tab.~2). 

First, we see that the fitted stubbornness rate is usually much lower when juries are hung ($\hat{f}<1$), which suggests that, under hung conditions, jurors significantly reduce the rate at which they stick to their current opinion. Also, the positive estimated values of $\hat{\alpha}$ indicate that juries are more likely to stop deliberating when they reach near-consensus. Further, $\hat{p}>0.5$ implies herding occurs within the jury, and $\hat{\mu}>0$ $\text{min}^{-1}$ implies jurors keep their most recent opinion with increasing stubbornness. 

\begin{figure}[t]
	\hspace{-12pt}
	\centering
	\includegraphics[width=0.99\columnwidth]{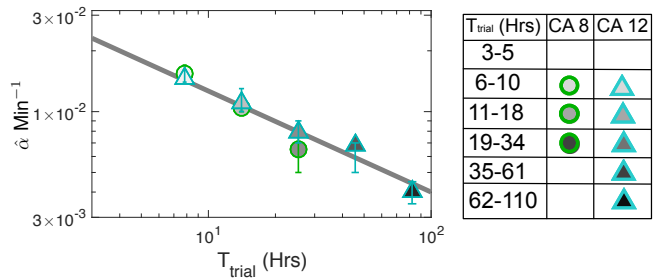}
	\caption{The scaling of the stopping rate, $\hat{\alpha}\sim\hat{q}$ (see Supporting Information), versus $T_\text{trial}$ showing that $\alpha\sim(T_\text{trial})^{-1/2}$ across CA, WA, and NE datasets. Error bars represent 90\% confidence intervals in the mean.
	}
	\label{fig:AlphaVsTtrial}
\end{figure}

In Fig. \ref{fig:AlphaVsTtrial}, we see that a parameter in the influence model, $\alpha$, follows the power law relation $\alpha\sim(T_\text{trial})^{-1/2}$, which agrees with Fig. \ref{fig:JuryData}b because $T_\text{delib}\sim\alpha^{-1}$ (Eq. \eqref{eq:q}). We propose a possible mechanism for the scaling relationship $T_\text{delib}\sim(T_\text{trial})^{1/2}$ : over the course of a trial, the amount of data juries will deliberate on, $D$, might follow a random walk with a reflecting boundary condition at 0, which implies that $\hat{\alpha}^{-1}\sim\langle T_\text{delib}\rangle\sim\langle D\rangle\sim(T_\text{trial})^{1/2}$ (see Supporting Information).

We also notice that, across all the data, the herding probability, $p$, is highest when juries are smallest (Tabs. 1 \& 2), while this value drops significantly for datasets with larger $N$ (p-value $<0.05$ between any $N=6$ dataset and any $N=12$ dataset). Previous studies on jury size \cite{MetaAnalysisJurySize1}, found that larger juries become hung more frequently, possibly because they have a minority opinion able to better resist the majority. Our study provides evidence of this explanation because larger juries have smaller $p$ values, and therefore jurors that are less likely to follow the majority opinion. We should caution, however, that influence is not necessarily homogenous across jurors, which may affect our results. 

\section*{Discussion}

We find that models in which jurors make decisions independent from each other disagree with the data. On the other hand, models in which jurors are influenced by each other agree well, at least qualitatively, with the data. Importantly, we found best agreement from a model in which jurors display tendencies to both follow one another and also increasingly stick to their current opinion. This type of behavior was also previously found to be important for explaining voting patterns in elections \cite{CombineModels}, which suggests that it may be a fundamental mechanism of group decision-making. 

Future work is necessary to better understand whether stubbornness or influence can hurt or help collective wisdom. In a recent theoretical paper \cite{JuryThmCorrel}, correlations between jurors were found to sometimes create judgments with lower accuracy than individual jurors when they need to reach a simple majority. In contrast, sequential voting, in which individuals base their decision on the popularity of decisions in the past, has been shown to significantly improve the wisdom of crowds \cite{Celis2016,Krafft2016}. We are not aware of any paper that discusses how stubbornness can empirically help or hurt deliberation, nor does our research directly address how stubbornness and/or influence affects the quality of jury decisions.

Our work could also be extended by building more accurate models and better addressing data heterogeneity. Most of the data is statistically significantly different from the model, based on the two-dimensional Kolmogorov-Smirnov test (p-value $< 0.1$) \cite{NumericalRecipes}, pointing to a need for more nuanced models to better explain the data. Another, more fundamental problem, however, in the datasets is heterogeneity: trials vary in complexity and jurors differ across trials, which can affect how decisions are reached. This may be addressed, however, with controlled experiments in which several groups separately deliberate on the same, or very similar, information. Data on how opinions change over time, as well as the time for juries to reach a verdict can provide tantalizing clues about the underlying mechanism of opinion dynamics.

\section*{Methods and Materials}

\subsection*{Gathering Data}
The jury data we study is taken from Multnomah County, Oregon \cite{ORJuryData}, San Francisco County, California \cite{CAJuryData}, Thurston County, Washington, and Douglas County, Nebraska \cite{USJuryPaper1,USJuryDataPaper2}. In the CA and OR datasets, the deliberation time and final vote are known, which can affect each other, but the OR dataset, unlike the other datasets, does no record $T_\text{trial}$. The CA dataset bins $T_\text{trial}$ in days, but the WA and NE datasets record both hours and days (roughly 4.5 hours per day in court), therefore we convert each trial day in the CA data into 4 hours. We removed all data where we did not have both the trial time, deliberation time in hours, and final vote in the CA data. Furthermore, we focus on trials in which jurors only vote on one count to simplify our study (this only removes 138 trials total) and the OR dataset only records the most important count if multiple exist \cite{ORJuryData}. Once cleaned, we have $53$ trials for CA 6, $338$ trials for CA 8, and $1726$ trials for CA 12 out of 6482 total trials. We do not know whether the kept data was unknowingly biased, although the qualitative similarities suggest that any bias should not significantly affect our results (Fig. \ref{fig:JuryData}). We also removed all data where we did not simultaneously know deliberation time and final vote in OR data (only 4 trials were removed; once cleaned, there were $207$ trials for OR 6 jury data, and $951$ trials for OR 12 jury data). Finally, we removed data where we did not simultaneously know both the trial time and deliberation time in the WA and NE data. This removed 10 trials for the WA dataset, and 21 trials for the NE dataset (in the cleaned data, there were $141$ and $135$ trials, respectively). 
All mean confidence intervals in the data come from bootstrapping data $10^4$ times. 

\subsection*{Fitting The Data}

In the OR dataset, some trials are criminal trials, which have different rules about when juries are hung (see Fig. \ref{fig:ORAttributes} \& Tab.~1) \cite{OregonHungJury}, therefore we primarily focus on civil trials. In the CA dataset, all trials were civil trials. In the WA and NE datasets, on the other hand, the final vote was not recorded, therefore we did not attempt to model the dynamics.

In the influence model, juror opinions are initially binomially distributed, with each juror having a probability $b$ of an initially for-plaintiff opinion. The parameter $b$ was chosen such that the probability of simulated juries initially voting for the plaintiff plus $1/2$ the probability of simulated juries being evenly split was equal to $\textrm{Pr}(V^\text{f}_\text{p} > N/2)$ in the dataset. This ensured that the final distribution had a similar value for $\textrm{Pr}(V^\text{f}_\text{p} > N/2)$. 
In addition, we somewhat arbitrarily set the timestep in simulations to be 1 minute, but simulations with significantly smaller or larger timesteps (as small as 15 seconds, or as large as 4 minutes) are not usually statistically significantly different (p-value$>0.1$ using the likelihood ratio test \cite{CSNPowerLaw}). An exception to this rule is CA12 with $T_\text{trial}=6-10$ and $11-18$ hours, where the timesteps of 15  seconds and 1 minute are not statistically different, but both are preferred over timesteps of 4 minutes (p-values vary between $0.006$ and $0.09$). Furthermore, $q_0$ is arbitrarily set to $0.3 \hat{\alpha}$, but varying this value between $0.1 \hat{\alpha}-1.0\hat{\alpha}$ similarly produces statistically equivalent fits (p-values $>0.1$). We cannot set this value to 0, however, because it would mean juries never stop deliberating when they are evenly split, which is in disagreement with the data.

To find $\hat{p}$, $\hat{\alpha}$, $\hat{\mu}$, and $\hat{f}$, we use maximum likelihood estimation, and then use the log-likelihood function to compare the quality of fits. Some values were predicted to be nearly 0 in the model, even though they existed in the data, therefore, we added a small base probability of between $10^{-4}$ and $10^{-14}$ to the models with no significant qualitative changes (all values shown are with a base probability of $10^{-11}$). 
Finally, the distributions we used to fit the influence models to the data were created from $1.6\times10^5$ simulations per parameter value. There was an inherent limit in the probability resolution ($6\times10^{-6}$), but we do not believe this significantly affects our results. All parameter confidence intervals come from bootstrapping and fitting the data $10^4$ times. 

\subsection*{Data Archival}
All data is publicly available from the following references: \cite{ORJuryData,CAJuryData,USJuryPaper1}.

\subsection*{Acknowledgements}
Our work is supported by the Army Research Office under contract W911NF-15-1-0142. KB would like to thank Nicholas Pace and Walter Fontana for enlightening discussions.

\section*{Supporting Information (SI)}

\subsection*{How Deliberation Time is Affected by the Final Vote}
It might not be intuitive why the deliberation time, $T_\text{delib}$, is highest when jurors are near consensus in both the data and the influence model, while $T_\text{delib}$ is lower when jurors are evenly split (see Figs.~\ref{fig:JuryData} \& \ref{fig:JuryModels} in the main text). In this section, we present a simple Markov chain model to better understand this finding. 

\begin{figure}[b]
	\hspace{-12pt}
	\centering
	\includegraphics[width=0.99\columnwidth]{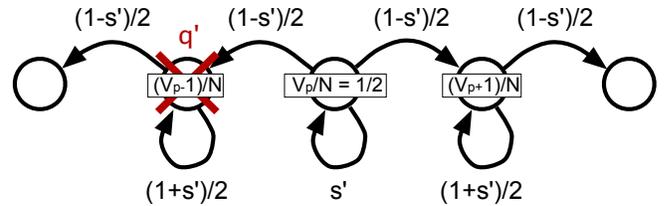}
\caption{A Markov model that qualitatively describes the dynamics of the influence model seen in the main text. We assume states change as a Markov chain, therefore with probability $s'$ we remain in state $V_\text{p}(t)/N=1/2$, but transition to $V_\text{p}(t)/N=1/2\pm1/N$ with probability $\frac{1-s'}{2}$. Once we are at a new state, we either transition to $V_\text{p}(t)/N=1/2\pm2/N$ with the same probability or stay in the current state. Finally, with probability $q'$, juries stop deliberating.}
\label{fig:SimpleModelSchematic}
\end{figure}

We find that the fraction of jurors voting for the plaintiff, 
$V^\text{f}_\text{p}/N = 1/2$ is rare, as seen in Fig.~\ref{fig:JuryModels}a in the main text. This finding is likely related to the influence model becoming the Majority Voter Model (MVM) when $\alpha=\mu=0$, i.e., when juries do not stop deliberating and there is no jury stubbornness, because $V_\text{p}(t)/N=1/2$ is known to be unstable past a critical point in the MVM when the influence of neighbors changes from weak (and opinions are evenly split) to strong (and there is near-unanimous agreement) \cite{MVMLattice,MVM}. Using this numerical finding, we create a similar, but much simpler, model in which the number of jurors voting for the plaintiff is represented as a node in a Markov chain, and there is a bias for juries to have greater agreement (see Fig. \ref{fig:SimpleModelSchematic}).

In the model, juries begin evenly split ($V_\text{p}(0)/N=1/2$) but can transition to a new state, $V_\text{p}(1)/N\pm 1/N$, with probability $(1-s')/2$. Once jurors reach this new state, they can achieve greater consensus, $V_\text{p}(1)/N\pm 2/N$, with probability $(1-s')/2$, or stay in the current state. This pattern can continue until juries stop deliberating with probability $q'$ at each timestep. 

Recall that, in the influence model seen in the main text, a juror will choose not to re-evaluate their opinion with probability $s$, and even if they do re-evaluate, they may choose to keep their original opinion, therefore it is reasonable for self-loops to exist in the Markov chain model. That said, because $s$ is often less than $1$, and $p>1/2$, it is reasonable to assume that opinions develop stronger pluralities over time, ergo the Markov chain model captures many qualitative features of the influence model.

Starting from time $t=1$, we find that the probability a jury is evenly split by the time they stop deliberating at time $t$ is
\begin{equation}
Pr(V^\text{f}_\text{p}/N=1/2,t) = q' (1-q')^{t-1} (s')^t
\end{equation} 
which implies that
\begin{equation}
Pr(1/2) = \frac{q' s'}{1-s' (1-q')}
\end{equation} 
and the probability the jury stops deliberating with an opinion $V^\text{f}_\text{p}/N=1/2 +1/N$ (or equivalently $V^\text{f}_\text{p}/N=1/2 -1/N$) at time $t$ is
\begin{widetext}
\begin{equation}
Pr(V^\text{f}_\text{p}/N=1/2+1/N,t) =Pr(1/2-1/N,t) =q' (1-q')^{t-1} \left\{\left[\frac{1+s'}{2}\right]^{t-1}-(s')^{t-1}\right\},
\end{equation}
\end{widetext}
and the probability over all time is
\begin{equation}
Pr(1/2\pm1/N) =\frac{q'(1-s')}{[(q'-1) s'+1] [(q'-1) s'+q'+1]},
\end{equation}
where we use $\pm$ to emphasize that the probabilities for $V^\text{f}_\text{p}/N=1/2+1/N$ and $V^\text{f}_\text{p}/N=1/2-1/N$ are the same. Using $Pr(1/2,t)$ and $Pr(1/2\pm1/N,t)$, we can also find the mean deliberation time conditioned on the final vote:

\begin{equation}
\langle T_\text{delib}(
1/2)\rangle 
 = \frac{s' q'}{[1-s'(1-q')]^2}
\end{equation} 
and
\begin{widetext}
\begin{equation}
\langle T_\text{delib}(1/2\pm1/N)\rangle =\frac{q' \left\{(q'-1)^2 s'^3-[(q'-2) q'+3] s'+2\right\}}{[(q'-1) s'+1]^2 [(q'-1) s'+q'+1]^2},
\end{equation}
\end{widetext}
where $\langle .\rangle$ is the average. If $s'\rightarrow 0$ (in other words, $V_\text{p}(t)/N=1/2$ is very unstable), then we find that 
\begin{equation}
Pr(1/2) = s'q',
\end{equation} 
and
\begin{equation}
Pr(1/2\pm1/N) = \frac{q'}{q'+1},
\end{equation}
In comparison
\begin{equation}
\langle T_\text{delib}(1/2)\rangle = q' s',
\end{equation}
and
\begin{equation}
\langle T_\text{delib}(1/2\pm1/N)\rangle =\frac{2 q'}{(q'+1)^2}.
\end{equation}
The probability that deliberation stops at $V^\text{f}_\text{p}/N=1/2$ is small, but so is the time that this deliberation would subsequently take. In comparison, $V^\text{f}_\text{p}/N=1/2\pm1/N$ is more likely, but mean deliberation time is subsequently higher. If we continue to $V^\text{f}_\text{p}/N=1/2\pm2/N$, $T_\text{delib}$ is expected to further increase because it takes at a minimum number of timesteps to reach the state. In short, the Markov chain model helps explain why $T_\text{delib}$ is low when the jury is evenly split, even though the probability for a jury to be evenly split is low as well. Furthermore, the Markov chain model helps explain why deliberation increases with greater consensus, at least until $V^\text{f}_\text{p}/N\approx0.3$ and $V^\text{f}_\text{p}/N\approx0.6$, when quitting rates substantially increase in the influence model, therefore lowering $T_\text{delib}$ again.

\subsection*{Random Walk Stopping Rate}
If we assume that the amount of information jurors accumulate is $D$, which we assume follows a random walk with a reflective boundary at $D=0$ (people cannot have negative information), and the amount of time users deliberate scales as $T_\text{delib}\sim D$, then $\alpha^{-1}\sim D$, where $\alpha$ is proportional to the quitting rate in the influence model. 
To better understand how $D$ affects the dynamics, recall that 

\begin{equation}
Pr(D|T) = \binom{T+1}{(T + D - 1)/2} \frac{D}{T + 1},
\end{equation}
where $T$ is the number of timesteps. Taking $T$ to be large, using the Sterling's formula, and dropping non-leading terms,

\begin{equation}
Pr(D|T) \approx \frac{2 D}{T} e^{-(D^2/ T)}.
\end{equation}
This immediately implies that $\langle D\rangle \sim T^{1/2}$. $T$ is not, as of yet, explicitly defined, because $T$ is still the number of timesteps and not an actual time. We can, however, set $T\sim T_\text{trial}$, and, because $T_\text{delib}\sim D$, $T_\text{delib}\sim T^{1/2}\sim T_\text{trial}^{1/2}$, in agreement with what we find empirically, therefore $\hat{\alpha}\sim T_\text{trial}^{-1/2}$. 

\begin{figure}[t]
	\centering
	\includegraphics[width=0.99\columnwidth]{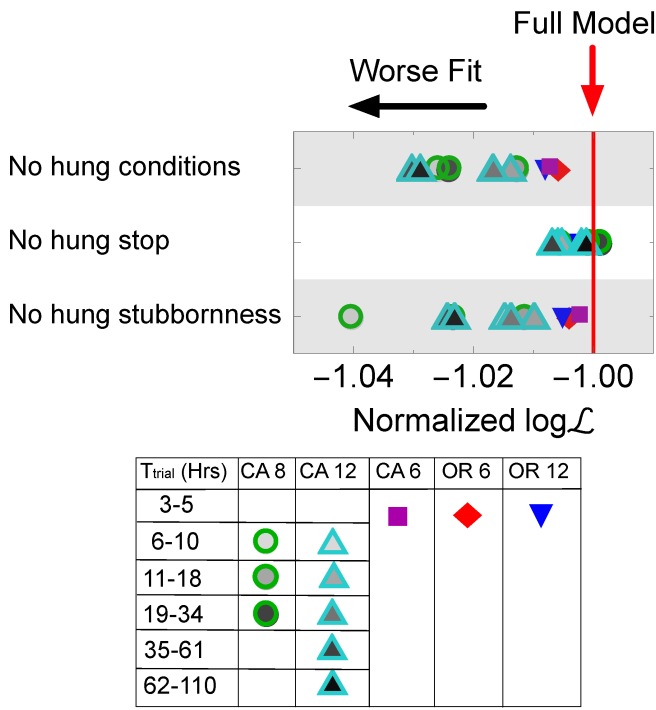}
	\caption{\label{fig:NoHungComparison}A comparison between the normalized log-likelihood functions of three model variants. $-1$ corresponds to the full model, and the more negative the log-likelihood, the worse the fit. The no hung conditions model creates consistently poorer fits than the full model, but if $q$ is independent of whether the jury is hung and  $\mu_\text{eff}(t)$ is Eq.~\eqref{eq:mu} in the main text (``No hung stop''),  the fit is very close to the full model, although this assumes, unnaturally, that juries with evenly split verdicts never stop deliberating. Instead, when we let $\mu_\text{eff}(t)=\mu$ (``No hung stubbornness''), the fit is roughly as poor as the model without any hung conditions. 
	}
\end{figure}
\subsection*{Alternative Jury Models}
We mention in the main text that removing all hung conditions in the herding and stubbornness model will produce a poorer fit (see Fig.~\ref{fig:ModelComparisons} in the main text and Fig.~\ref{fig:NoHungComparison}). To better understand why this is the case, we separately remove the dependence of the quitting rate, $q$, and stubbornness rate, $\mu_\text{eff}(t)$, on whether the jury is hung. In the former case, we see a small change in the likelihood function, but in the latter case, the likelihood function has a more significant drop. This suggests that jurors depend more on changing their stubbornness rate than changing their quitting rate when they avoid hanging.

\subsection*{Oregon Criminal Cases}
In this section, we compare data and fits for criminal and civil cases in Oregon. The reason we separate the data is both because the requirements for a verdict are different (ten out of twelve jurors are need to agree instead of nine out of twelve, although five out of six still need to agree in six-person juries), and the motivations for reaching a decision may be different. Overall, we find quantitatively similar findings between criminal and civil cases. 

\begin{figure}[tb]
	\centering
	\includegraphics[width=0.99\columnwidth]{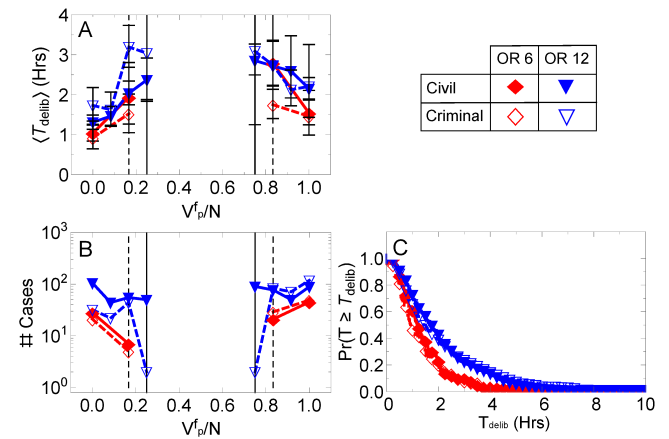}
\caption{\label{fig:ORAttributes}
{OR 6 and OR 12 Data attributes for Criminal and Civil Cases.} 
 (a) The mean deliberation time peaks when the $V^\text{f}_\text{p}/N\approx0.3$ and $0.6$, where $V^\text{f}_\text{p}/N$ is the fraction of jurors voting for the plaintiff (or voting guilty in criminal cases). (b) The distribution of $V^\text{f}_\text{p}/N$ shows peaks when $V^\text{f}_\text{p}/N\ge \phi$ or $V^\text{f}_\text{p}/N\le 1-\phi$, where $\phi$ is the thresholds between which juries hang ($\phi=0.75$ for civil cases and $0.833$ for criminal cases) \cite{ORJuryData,CAHungJuries}. (c) The complimentary cumulative distribution of deliberation times is heavy-tailed across datasets. Data is taken from \cite{ORJuryData} and error bars represent 90\% confidence intervals in the mean.}
\end{figure}

First, we compare OR 6 and OR 12 attributes seen in Fig. \ref{fig:JuryModels} of the main text (Fig. \ref{fig:ORAttributes}). We find that $\langle T_\text{delib}\rangle$ is higher for OR 6 criminal cases compared to civil cases, but the trend is not as clear for the OR 12 cases (Fig. \ref{fig:ORAttributes}a). That said, in all cases we see that $\langle T_\text{delib}\rangle$ is higher when there is greater disagreement among jurors. We also find that juries are commonly found to reach a verdict and hung juries are rare (Fig. \ref{fig:ORAttributes}b). Finally, we see that $Pr(T_\text{delib})$ is almost exactly the same for both civil and criminal cases. 

\begin{table*}[t]
\centering
Table 2: Model Parameter Fits: OR 6 \& OR 12 Criminal Cases
\label{tab:ModelFitsOR}
\vspace{1em}
\begin{minipage}{.98\textwidth}
\centering
{\renewcommand{\arraystretch}{0.95}
\setlength\tabcolsep{1pt}
\begin{tabular}{|>{\columncolor{Gray}}r| c| c| >{\columncolor{Red}}c| >{\columncolor{Red}}c| >{\columncolor{Orange}}c|>{\columncolor{Orange}} c| >{\columncolor{Green}} c|>{\columncolor{Green}}  c|>{\columncolor{LightCyan}} c|>{\columncolor{LightCyan}}  c|}
  \hline
  \multicolumn{11}{|c|} {\cW \bf Parameter Fits \bigstrut} \\ \hline 
  Data & $T_\text{trial}$ (Hrs) & Num. Trials & $\hat{p}$ & 90\% Conf. & $\hat{\alpha}$ ($\text{min}^{-1}$)& 90\% Conf. & $\hat{\mu}$ ($\text{min}^{-1}$) & 90\% Conf. & $\hat{f}$ & 90\% Conf. \\
\hline 
\vspace{-7pt}
&&&&&&&&&&\\
OR 6 Crim. & -- &104&0.961 & 0.92-1.0   & 0.0181 &0.016-0.022   & 0.36&0.25-0.50 & 0.045& 0.0-0.1\\
OR 12 Crim. & --&374&0.904 & 0.89-0.91& 0.0116 & 0.010-0.014 & 0.018  &0.01-0.05 & 0.32& 0.0-0.6 \\
\hline
\end{tabular}
}
\end{minipage}
\end{table*}

\begin{figure}[t]
	\centering
	\includegraphics[width=0.99\columnwidth]{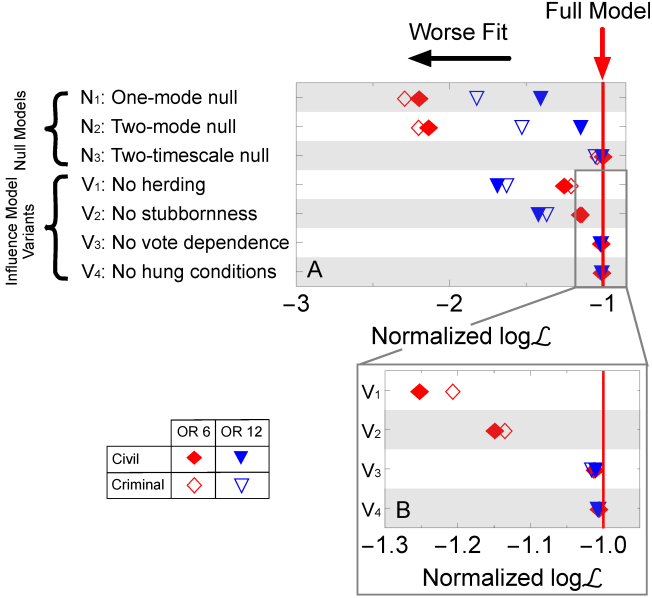}
\caption{\label{fig:ORFitComparison}
Comparison of Models for OR 6 and OR 12 civil and criminal cases. Normalized log-likelihood functions for the null models and the influence model variants to illustrate comparison with the full influence model. For each dataset indicated in the legend, log-likelihood functions for these models were normalized by $|\text{log}(\mathcal{L}_\textrm{full})|$, the log-likelihood function of the full model, therefore models above -1 explain the data better than the full influence model, while those below -1 explain the data worse. (a) The relative fit of the one-mode, two-mode, and two-timescale null models, along with the ``no herding'' model, in which $p=0.5$, ``no stubbornness'' model with $\mu=0$, ``no vote dependence'' model, in which the model dynamics do not depend on the number of jurors voting for the plaintiff (or voting guilty in criminal cases), and the ``no hung conditions'' model, in which jury dynamics do not depend on whether the jury is currently hung. (b) In a zoomed-in graph, the influence model variants seen in (a) perform worse than the full model.}
\end{figure}

\begin{figure}[tbh]
	\centering
	\includegraphics[width=0.99\columnwidth]{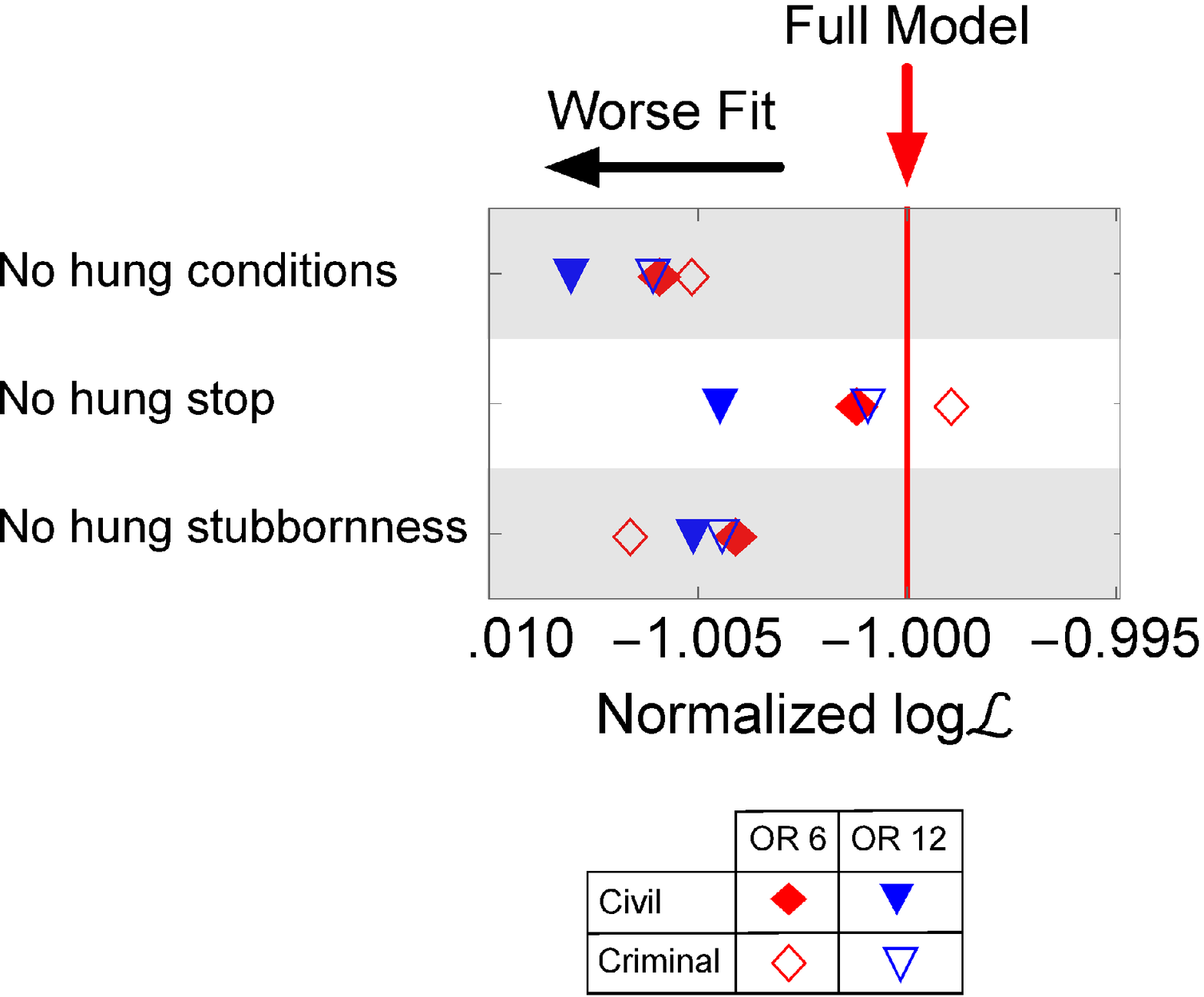}
	\caption{
 \label{fig:ORNoHungComparison}A comparison between the normalized log-likelihood functions of three model variants for OR 6 and OR 12 civil and criminal data. Compare to Fig. \ref{fig:NoHungComparison}.
}
\end{figure}

Next we compare the fits for civil and criminal cases. Overall, we find that civil and criminal cases fit each model similarly well (Fig. \ref{fig:ORFitComparison}). For example, the one-mode and two-mode null models give some of the worst fits, and the two-timescale model was the best null model, although it was still worse than the full influence model. Furthermore, removing either herding or stubbornness from the full influence model produces a much worse fit, while removing the vote dependence or hung conditions has a much smaller effect. We also see the same qualitative trends when we separately remove the dependence of the stubbornness rate or quitting rate on whether the jury is hung (Fig. \ref{fig:ORNoHungComparison}). In both the criminal and civil cases, removing the stubbornness rate's dependence on whether a jury is hung creates a significantly worse fit compared to removing the quitting rate's dependence. 

\subsection*{Correlations Between Jury Attributes}

\begin{figure}[h]
	\centering
	\includegraphics[width=0.85\columnwidth]{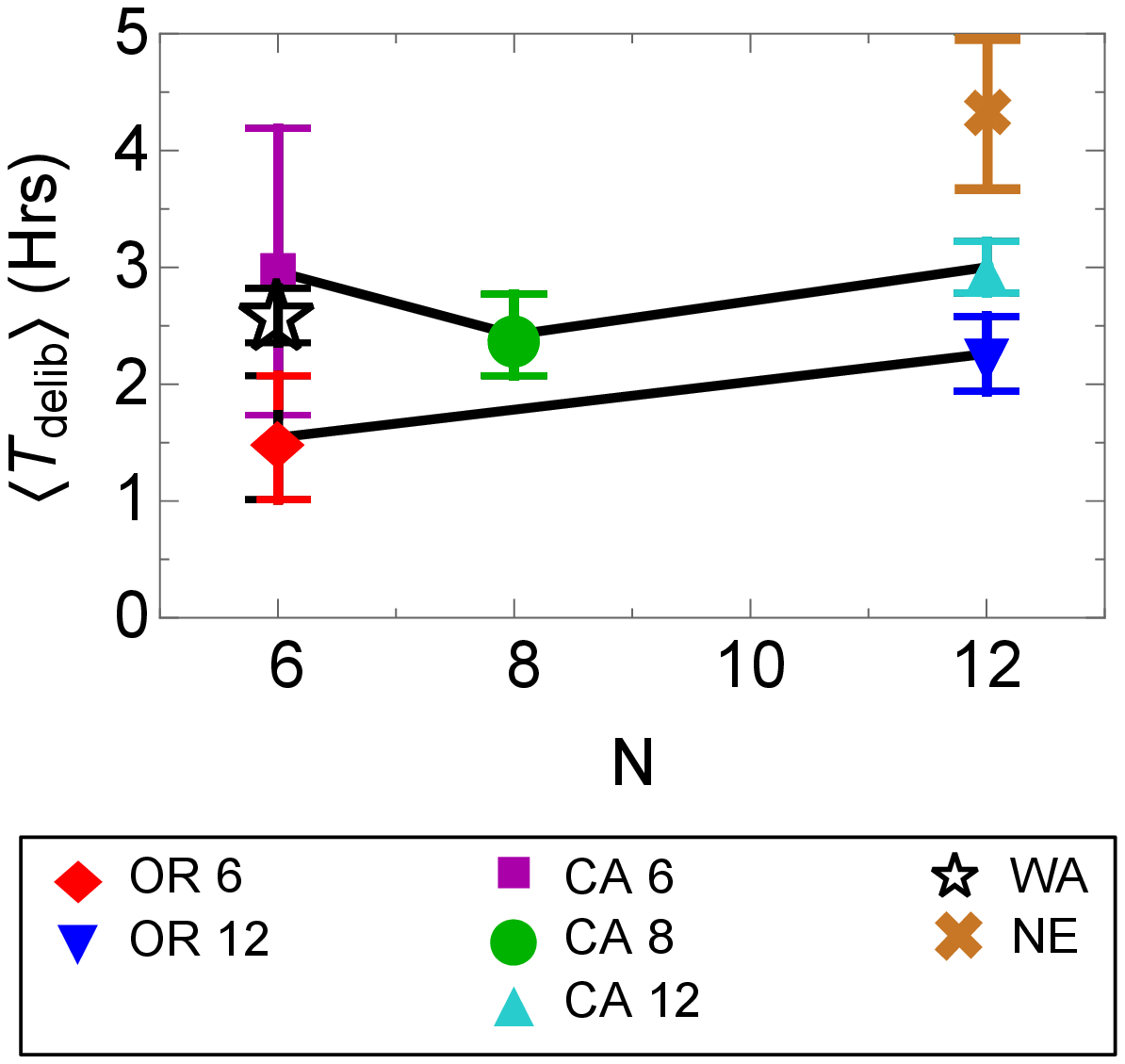}
	\caption{\label{fig:TdelibVsN}The mean deliberation time, $T_\text{delib}$ versus number of jurors for all the data sets. Error bars represent 90\% confidence intervals in the mean (via bootstrapping).
	}
\end{figure}

\begin{figure}[h]
	\centering
	\includegraphics[width=1\columnwidth]{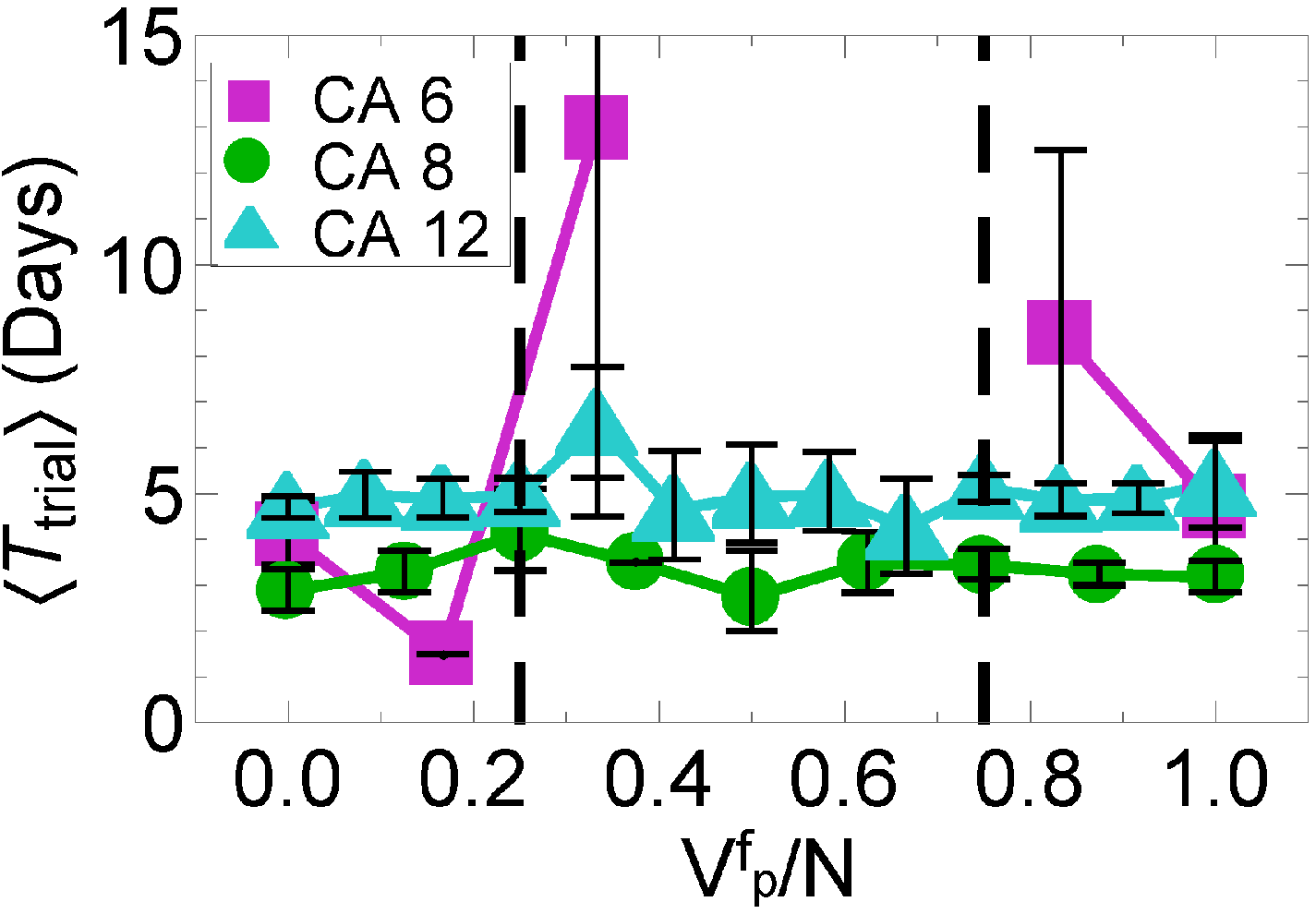}
	\caption{\label{fig:TrialTimeVsFinalVote}The mean trial time, $T_\text{trial}$ versus the final fraction of jurors voting for the plaintiff, $V^\text{f}_\text{p}/N$, for CA 6, CA 8, and CA 12. Error bars represent 90\% confidence intervals in the mean (via bootstrapping). We find that the trial time does not change significantly, which suggests the dependence of $T_\text{delib}$ on $V^\text{f}_\text{p}/N$ (Fig.~\ref{fig:JuryData}a in the main text), and the scaling between the deliberation time, $T_\text{delib}$ and $T_\text{trial}$ (Fig.~\ref{fig:JuryData}b in the main text) are mostly independent effects.
	}
\end{figure}

In this section, we discuss correlations between various attributes, in order to better understand how to model jury dynamics. First, we look at how the jury size affects the deliberation time (Fig. \ref{fig:TdelibVsN}), and notice very little correlation between the two. This contrasts with many models of opinion dynamics in which deliberation time strongly correlates with system size \cite{VMConsensus,MajorityRule}. Next, we compare how the trial time depends on the final vote (Fig. \ref{fig:TrialTimeVsFinalVote}). Interestingly, although both the trial time and the final vote strongly affect the deliberation time (Fig.~\ref{fig:JuryModels} in the main text), neither are strongly correlated with each other. We use this property to find separate mechanisms for the correlation between each attribute and deliberation time.

\begin{figure}[tb]
	\centering
	\includegraphics[width=0.99\columnwidth]{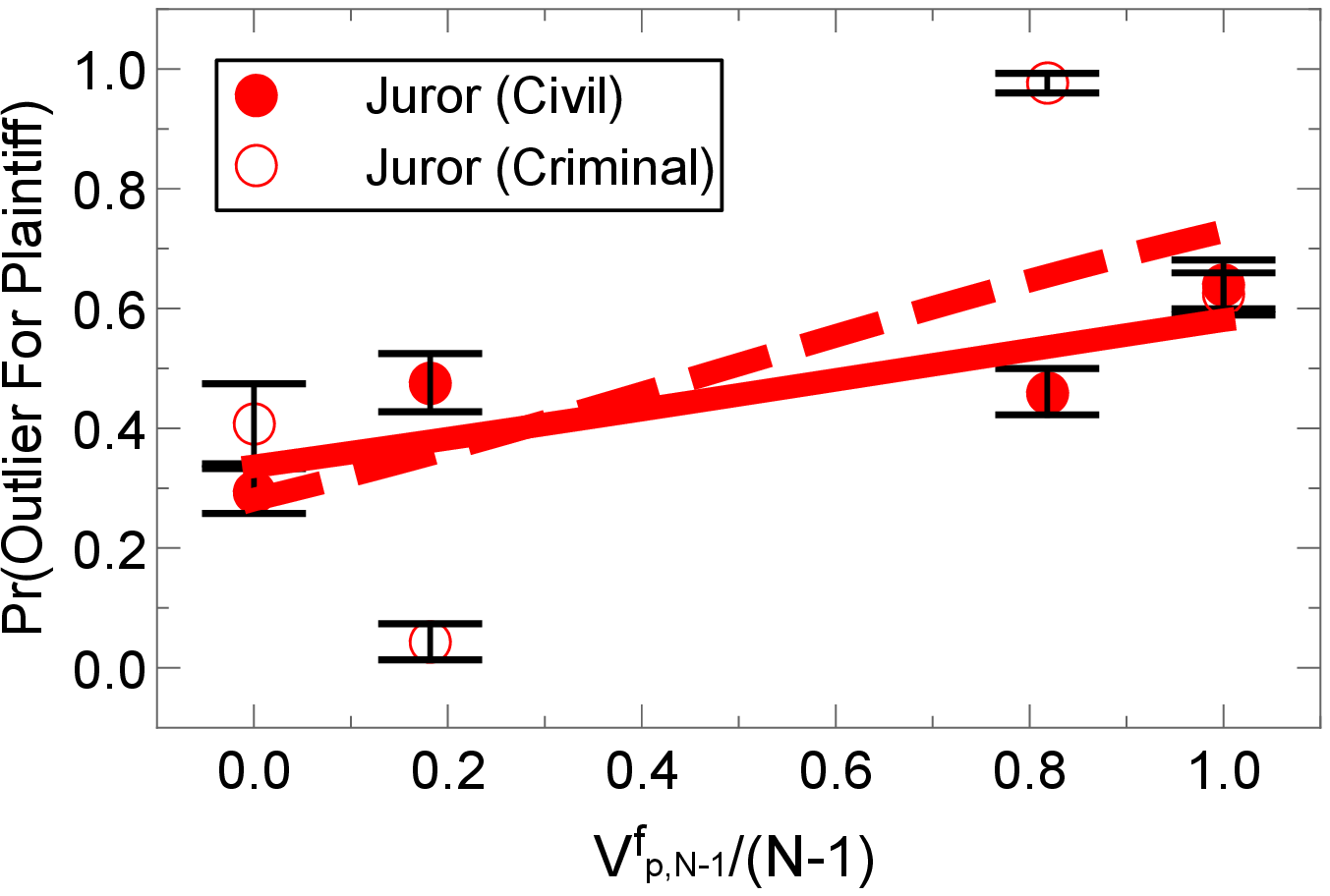}
	\caption{\label{fig:VoteCorrelation}The probability of a typical juror voting for the plaintiff (or guilty in criminal cases) versus the fraction of other jurors voting for the plaintiff ($V^\text{f}_{\text{p},N-1}/(N-1)$) for $N=12$ jurors in the OR 12 civil and criminal datasets. Fits are logistic regression, $y=\frac{1}{1+e^{-(\beta_0 + \beta_1 x)}}$, with coefficient $\beta_1=1.0$ (95\% confidence $[0.6-1.4]$) for civil cases, and $\beta_1=1.9$ (95\% confidence $[1.4-2.5]$) for criminal cases.
	}
\end{figure}

Finally, we plot the probability a typical voter will vote for the plaintiff (or vote guilty in criminal cases), $Pr(\text{Outlier For Plaintiff})$, versus the vote of all the other jurors for OR 12 (Fig. \ref{fig:VoteCorrelation}). We find a strong correlation between the two in civil cases and criminal cases, therefore juror opinions are not independent, which gives strong evidence that herding may exist in juries. We find $Pr(\text{Outlier For Plaintiff})$ by determining how many trials end with verdict $V^\text{f}_{\text{p}} = V^\text{f}_{\text{p},N-1}+1$, corresponding to the outlier juror voting for the plaintiff, and how many trials end with $V^\text{f}_{\text{p}} = V^\text{f}_{\text{p},N-1}$, corresponding to the outlier juror voting for the defendant. The probability, $Pr(\text{Outlier For Plaintiff})$, is simply

\begin{widetext}
\begin{equation}
Pr(\text{Outlier For Plaintiff}) = 
 \frac{Pr\left(V^\text{f}_{\text{p},N-1}+1\right)}{Pr\left(V^\text{f}_{\text{p},N-1}+1\right)+Pr\left(V^\text{f}_{\text{p},N-1}\right)}.
\end{equation}
\end{widetext}

\newpage
\bibliography{JuryDynamicsBib}
\end{document}